# A simple analytical thermal model to solar cavity receivers

Hamou Sadat

Institut Pprime, Université de Poitiers, 40 Avenue du Recteur Pineau 86022 Poitiers, France

**Abstract:** This article presents a closed form analytical solution to estimate solar receiver surface and fluid temperatures. An approximation and its domain of validity (in term of the value of a small parameter) are also proposed. These simple models are then applied to a large and a small cylindrical cavity. Finally, the model is applied to an experimental hemispherical coiled cavity.

## 1. Introduction

Concentrating solar thermal technology has been the subject of a great deal of research and development during the last decades [1-5]. When high temperatures were sought for, parabolic dish reflectors have been widely used [6-7]. Solar energy impinging on the reflector surface is concentrated in a small region around the focal point of the dish where a receiver is placed. It is generally admitted that cavity receivers have the better efficiency because of multireflexions of visible and infrared radiations. These receivers have therefore been widely studied both experimentally and theoretically [8-10]. In such cavities, concentrated solar power is absorbed, heat is exchanged by radiation, convection and conduction with the environment and a useful power is generally transmitted to a circulating heat transfer fluid (HTF). Several detailed mathematical models describing all the above mentioned phenomena have been proposed so far. Radiosity or Gebhart methods (describing surface radiation) can be coupled to Navier-Stokes equations to obtain very fine and accurate but time consuming numerical models [11-12]. Simpler models necessitating the solution of a limited number of non-linear equations have also been used [13-14]. However, it appears, to the best of our knowledge, that there is a lack of simple analytical solutions. Hence, we present in this work, a very simple one-equation model which can be useful in the first stage of conception of such solar cavities. This model only uses the wall cavity and the working fluid temperatures as variables. This paper starts up by describing the thermal balance equation of a solar cavity and its closed form analytical solution. Next, an approximate solution is deduced for small values of a design parameter ε that will be defined later. These different explicit expressions are then applied to calculate the temperatures of two receivers (a large one and a small one). It will be seen that while the complete analytical solution is necessary to calculate the first receiver temperatures, the simplest approximate model is sufficient to deal with the second one. Finally, in order to assess the accuracy of the model, it has been used to calculate the exit heat transfer fluid temperature in the case of an experimental hemispherical receiver. A relative error of 14% was obtained .

## 2. Mathematical model

Of concern here is a solar cavity receiver of any geometrical shape (cylindrical, conical, hemispherical) with a heat transfer fluid (HTF) flowing in an annular space as depicted in Figure 1. It is assumed that the internal wall is isothermal at temperature $T_c$ and that the fluid enters at a temperature $T_{fi}$ with a mass flow rate $\dot{m}$. In this model, the fluid shall be considered to be very well mixed and isothermal at temperature $T_f$. The receiver is at steady-state condition and its outer surface is well insulated. A total reflected power P enters the cavity which losses heat par radiation ($Q^r$) and natural convection ($Q^{conv}$) through its aperture and by forced convection ($Q^{fluid}$) to the working fluid through its internal surface $A_c$. Heat conduction through the walls of the well-insulated cavity is neglected.

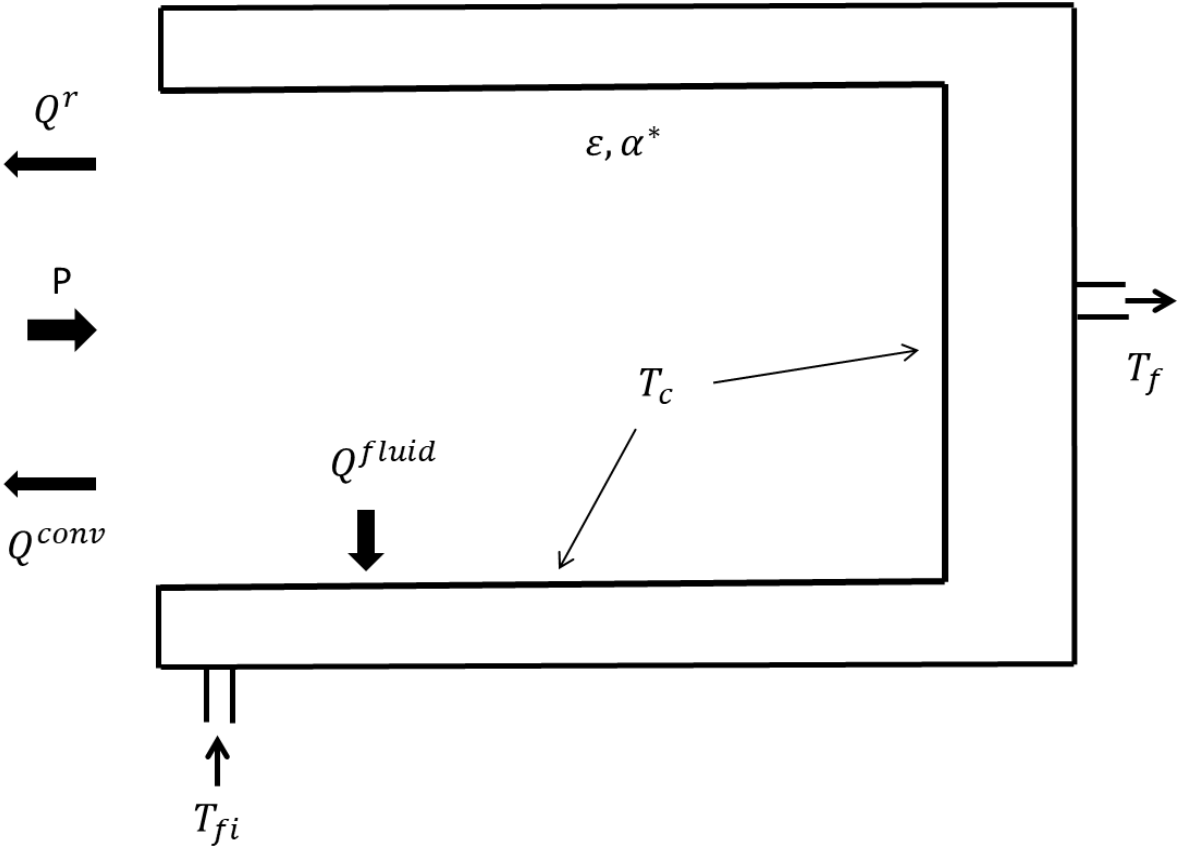

Figure 1: Schematic view of the solar cavity

The internal wall of the cavity is supposed to have different optical properties in the visible and long-waves radiations. Let $\alpha^*$, $\varepsilon_c$, $h_{cn}$ and $h_{cf}$ be the wall absorptivity in the visible, the infrared emissivity and the natural and forced convection exchange coefficients. An energy balance in the cavity can be written as follows:

$$\alpha^* P = \varepsilon_c A_c \sigma T_c^4 + h_{CN} A_c (T_c - T_{amb}) + h_{CF} A_c (T_c - T_f) \quad (1)$$

The irradiation E on the cavity wall shall be considered to be uniform ($E = P/A_c$) so that:

$$\alpha^* E = \varepsilon_c \sigma T_c^4 + h_{CN}(T_c - T_{amb}) + h_{CF}(T_c - T_f) \quad (2)$$

Introducing the fluid mass flow rate $\dot{m}$, its entrance temperature $T_{fi}$ and heat capacity $c_p$, the following equality holds:

$$h_{CF} A_c (T_c - T_f) = \dot{m} c_p (T_f - T_{fi}) \quad (3)$$

The fluid temperature is therefore simply related to the cavity temperature:

$$T_f = \frac{h_{CF} A_c}{(\dot{m} c_p + h_{CF} A_c)} T_c + \frac{\dot{m} c_p}{(\dot{m} c_p + h_{CF} A_c)} T_{fi} \quad (4)$$

Introducing (4) in (2), one obtains the following non linear equation:

$$a T_c^4 + b T_c - (\alpha^* E + c) = 0 \quad (5)$$

where:

$$a = \varepsilon_c \sigma; \quad b = \left(h_{CN} + h_{CF} \frac{\dot{m} c_p}{(\dot{m} c_p + h_{CF} A_c)}\right); \quad c = h_{CN} T_{amb} + h_{CF} \frac{\dot{m} c_p}{(\dot{m} c_p + h_{CF} A_c)} T_{fe} \quad (6)$$

Equation (5) can be solved numerically by using a Newton-Raphson or an iterative method. It can also be solved analytically by the Ferrari's method which gives the following cavity temperature:

$$T_c = \frac{1}{2}\left(\sqrt{\frac{2|q|}{\sqrt{t}} - t} - \sqrt{t}\right) \quad (7)$$

where: q=-b/a and r=$(\alpha^* E + c)/a$ , and $t = \sqrt{\sqrt[3]{\frac{q^2+\sqrt{q^4+4\left(\frac{4r}{3}\right)^3}}{2}} - \sqrt[3]{\frac{\sqrt{q^4+4\left(\frac{4r}{3}\right)^3}-q^2}{2}}}$ .

Equation (4) then gives the fluid temperature and the thermal efficiency of the receiver defined as the ratio between the power absorbed by the circulating fluid in the cavity and the solar power entering the receiver is deduced as follows:

$$\eta = \frac{\dot{m}c_p(T_f-T_{fi})}{P} \quad (8)$$

The analytical solution given by equation (7) is easy to calculate for any receiver cavity. However, the influence of the different parameters of the problem does not appear explicitly. We show in the next section that under some condition, a more explicit analytical expression can be derived.

## 3. Approximate explicit solution

Variable t can be written in the form:

$$t = \sqrt{\sqrt[3]{\frac{q^2+q^2\sqrt{1+\frac{4}{q^4}\left(\frac{4r}{3}\right)^3}}{2}} - \sqrt[3]{\frac{q^2\sqrt{1+\frac{4}{q^4}\left(\frac{4r}{3}\right)^3}-q^2}{2}}} \quad (9)$$

For small values of $\varepsilon = \frac{r^3}{q^4}$ , this can be approximated by:

$$t = \sqrt{\sqrt[3]{\frac{q^2+q^2(1+\frac{2}{q^4}\left(\frac{4r}{3}\right)^3)}{2}} - \sqrt[3]{\frac{q^2(1+\frac{2}{q^4}\left(\frac{4r}{3}\right)^3)-q^2}{2}}} \quad (10)$$

This leads to : $t = |q|^{\frac{2}{3}} - \frac{4r}{3|q|^{\frac{2}{3}}}$ and $\sqrt{t} = |q|^{\frac{1}{3}}\left(1 - \frac{2r}{3|q|^{\frac{4}{3}}}\right)$. Introducing these expressions in Equation (7), cavity temperature can be expressed as follows:

$$T_c = \frac{1}{2}\left(\sqrt{\frac{2|q|}{|q|^{\frac{1}{3}}\left(1-\frac{2r}{3|q|^{\frac{4}{3}}}\right)} - |q|^{\frac{2}{3}} + \frac{4r}{3|q|^{\frac{2}{3}}}} - |q|^{\frac{1}{3}}\left(1 - \frac{2r}{3|q|^{\frac{4}{3}}}\right)\right) \quad (11)$$

or:

$$T_c = \frac{1}{2}\left(\sqrt{|q|^{\frac{2}{3}}(1 + \frac{8r}{3|q|^{\frac{4}{3}}})} - |q|^{\frac{1}{3}}(1 - \frac{2r}{3|q|^{\frac{4}{3}}})\right) \quad (12)$$

One can now approximate the square root term and write:

$$T_c = \frac{1}{2}\left(|q|^{\frac{1}{3}}(1 + \frac{4r}{3|q|^{\frac{4}{3}}}) - |q|^{\frac{1}{3}}(1 - \frac{2r}{3|q|^{\frac{4}{3}}})\right) \quad (13)$$

This leads to the following expression of the cavity temperature:

$$T_c = \frac{1}{2}\left(|q|^{\frac{1}{3}}\left(\frac{6r}{3|q|^{\frac{4}{3}}}\right)\right) = \frac{r}{|q|} \quad (14)$$

Using the expression of $|q|$, one finally obtains the explicit form:

$$T_c = \frac{\alpha^* E}{h_{CN} + h_{CF}\frac{\dot{m}c_p}{(\dot{m}c_p + h_{CF}A_c)}} + \frac{h_{CN}T_{amb} + h_{CF}\frac{\dot{m}c_p}{(\dot{m}c_p + h_{CF}A_c)}T_{fi}}{h_{CN} + h_{CF}\frac{\dot{m}c_p}{(\dot{m}c_p + h_{CF}A_c)}} \quad (15)$$

Introducing the previous relation in Equation (4) leads to the explicit form of the fluid temperature :

$$T_f = \frac{h_{CF}A_c\alpha^* E}{\dot{m}c_p(h_{CN} + h_{CF}) + h_{CN}h_{CF}A_c} + \left[\frac{h_{CN}h_{CF}A_c}{\dot{m}c_p(h_{CN} + h_{CF}) + h_{CN}h_{CF}A_c}\right]T_{amb}$$
$$+ \left[1 + \frac{h_{CF}h_{CF}A_c}{\dot{m}c_p(h_{CN}+h_{CF}) + h_{CN}h_{CF}A_c}\right]\frac{\dot{m}c_p}{(\dot{m}c_p + h_{CF}A_c)}T_{fi} \quad (16)$$

When the HTF enters at the ambient temperature $(T_{amb} = T_{fi})$, this degenerates to:

$$T_f = \frac{h_{CF} A_c \alpha^* E}{\dot{m} c_p (h_{CN} + h_{CF}) + h_{CN} h_{CF} A_c} + T_{amb} \quad (17)$$

Finally, a further approximation can be obtained if: $h_{CN} \ll h_{CF} \frac{\dot{m} c_p}{(\dot{m} c_p + h_{CF} A_c)}$. It reads:

$$T_f = \frac{A_c \alpha^* E}{\dot{m} c_p} + T_{amb} \quad (18)$$

and $T_f$ is no longer depending on h_cf in this condition. Equation (18) is nothing but a thermal balance in which all the absorbed power is transferred to the circulating fluid.

## 4. Applications

This section is first devoted to the application of the previous models to the calculation of cavity and fluid temperatures of two hypothetical but still realistic solar receivers that can be modeled by one of the two analytical solutions given by equations () and (). Then, in order to assess the accuracy of the proposed simplified models, a comparison with experimental and numerical results obtained recently [15] are presented.

### 4.1 First example

The first example is a 2 m long by 2 m diameter cylindrical solar receiver receiving a total reflected power P=800 kW. A heat transfer fluid is circulating in lateral and back surfaces of the cavity. The other parameters are given in table 1. The forced convection exchange coefficient is difficult to estimate as it depends on the shape of the volume where the HTF is circulating and on the velocity. The cavity and fluid temperatures have therefore been calculated by using equations (7) and (4) for a forced convection exchange coefficient varying from 400 to 2300 W/m²K. Equation (5) has also been solved by an iterative method. The two approaches gave the same results which are presented on Figure 2. As expected the cavity temperature decreases (from 598K to 564K) with increasing exchange coefficient while the opposite is true for the fluid temperature which increases from 540K to 553K. The efficiency of this solar receiver increases slightly from 45.6% to 48% as it is shown on Figure 3.

| $h_{CN}$ (W/m²K) | $\dot{m}c_p$ (J/K) | $T_{amb}$ (K) | $T_{fi}$ (K) | $\alpha^*$ | $\varepsilon_c$ |
|---|---|---|---|---|---|
| 10 | 1520 | 300 | 300 | 0.6 | 0.6 |

Table 1 : parameters of example 1

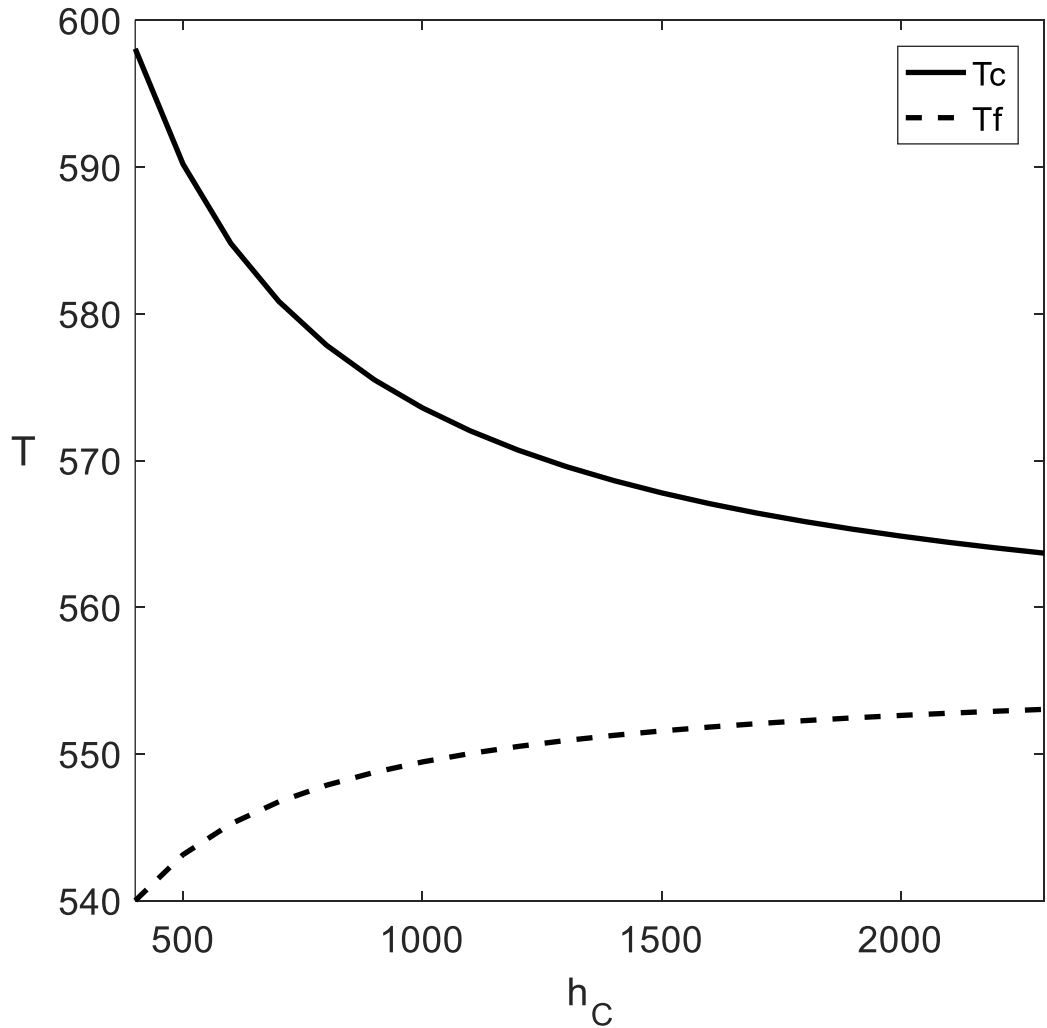

Figure 2: Cavity and fluid temperature versus heat exchange coefficient

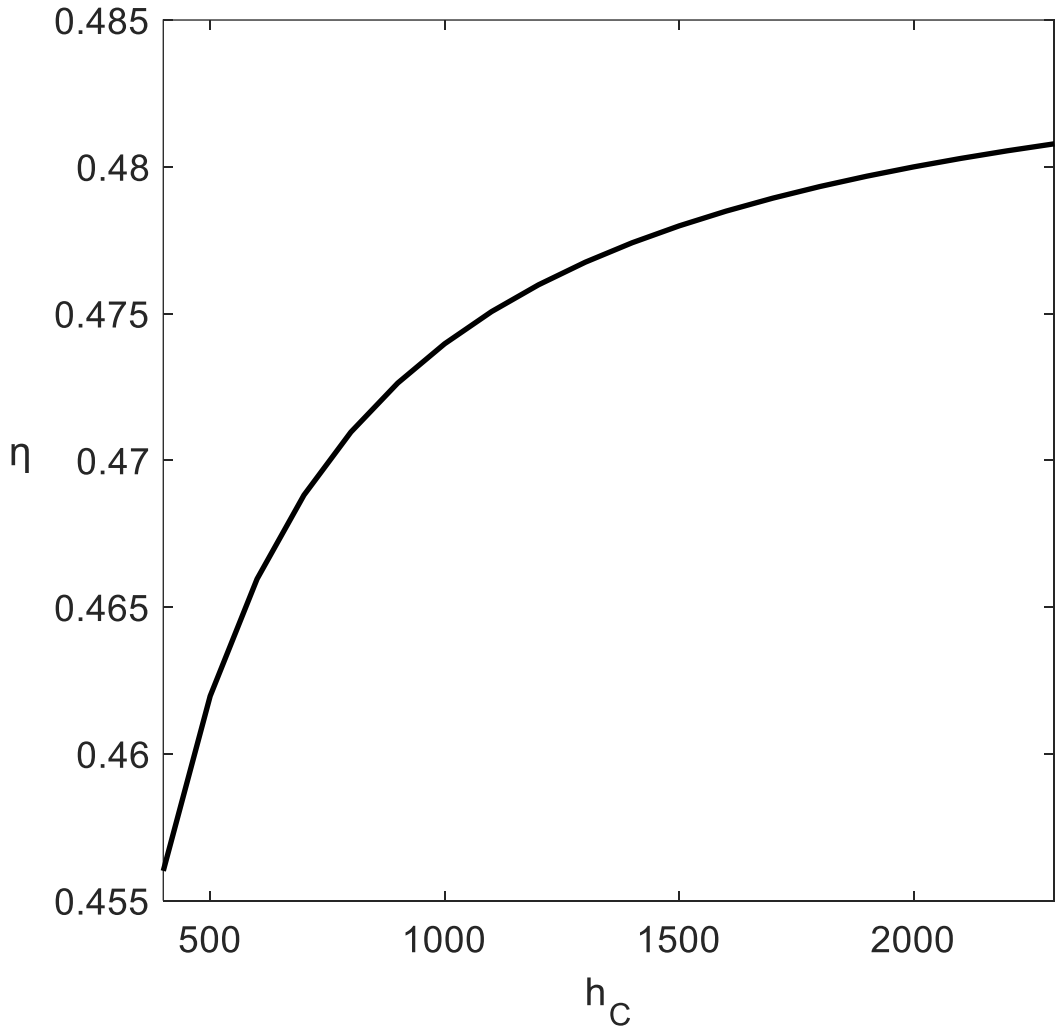

Figure 3: Efficiency of the receiver as a function of forced convection coefficient

**4.2 Second example**

As a second example, we consider herein a smaller receiver having a depth of 20 cm and inner diameter of 10 cm placed in the focal region of a parabolic dish with an aperture area of 2 m² under a solar irradiation of 1000 W/m². Water enters at an ambiant temperature equal to 294K and is circulating at a rate of 2.5 l/mn. The radiative properties were set to $\alpha^* = 0.6$ and $\varepsilon = 0.9$ and the solar irradiation on the cavity surface was equal to 28289 W/m². Here again, these operating conditions were used with different values of the forced convection heat exchange coefficient $h_{CF}$ ranging from 350 to 3300 W/m²K. Calculations of the cavity and fluid temperatures were carried out by using Equations (7) and (4). These temperatures are presented on Figure 4 as functions of $h_{CF}$. As seen previously, the cavity temperature decreases (from 344.65K to 305.85K) with increasing $h_{CF}$. However, the fluid temperature remains almost constant (between 300.53K and 300.9K). It is worth noting that these values are very close to the values given by equation (17) and ranging from 300.82K to 301.09K. They are also close to the constant fluid temperature (301.17K) obtained by using equation (18) with a difference comprised between 0.27K and 0.64K. Indeed, we are here in a case where the ratio ε is small (between $1.1 \cdot 10^{-3}$ and $6.6 \cdot 10^{-3}$) as highlighted on Figure 5 and where the condition: $h_{CN} \ll h_{CF} \frac{\dot{m}c_p}{(\dot{m}c_p + h_{CF}A_c)}$ also holds as the right hand term is ranging between 304.9 W/m²K and 1377.6 W/m²K. It can therefore be concluded that in the case of this small receiver, the two approximate explicit solutions are both of sufficient accuracy. Finally, the efficiency of this small receiver was found to increase from 54.59% to 57.75% with increasing $h_{CF}$.

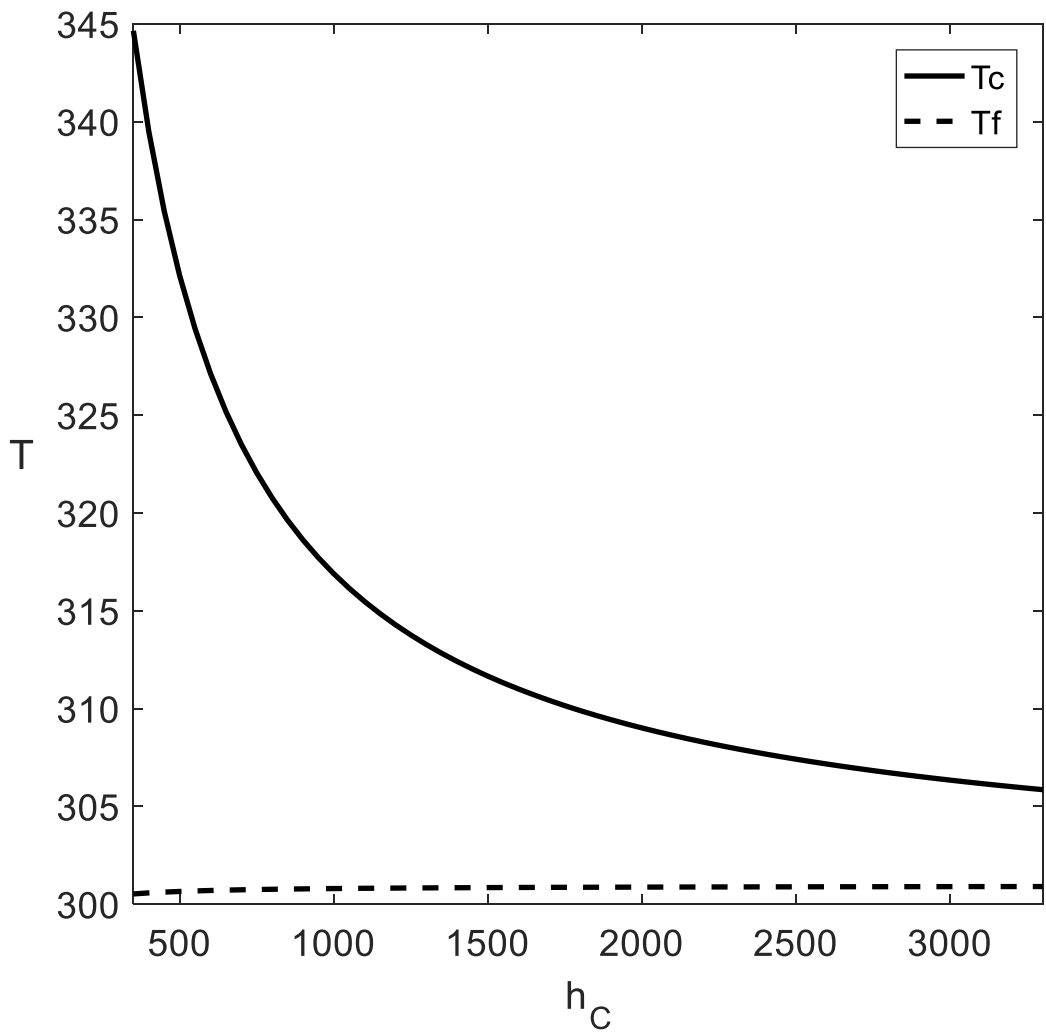

Figure 4: Cavity (bold line) and fluid (dashed line) temperatures as functions of forced convection coefficient

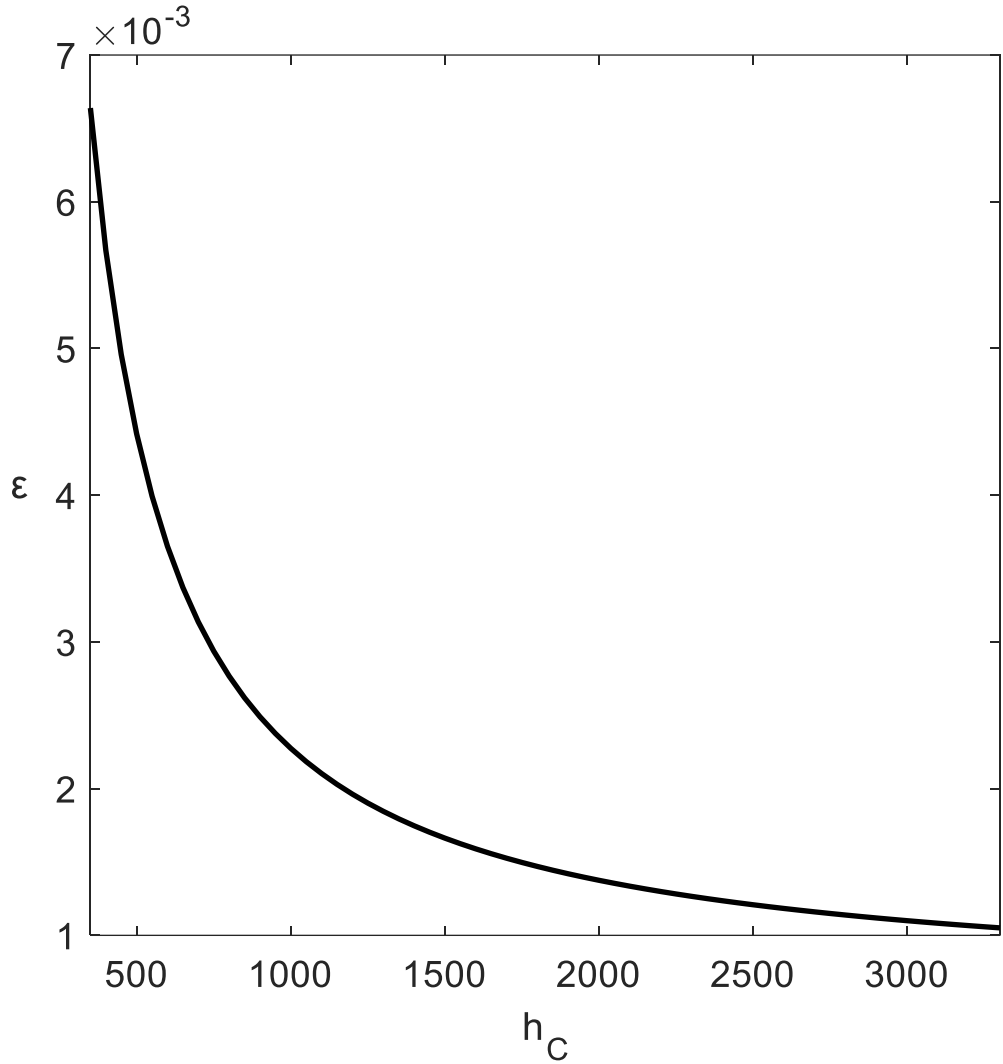

Figure 5: ratio ε as a function of forced convection coefficient

## 4.3 Comparison with experimental and numerical results

We now turn to the comparison of our simple analytical model with experimental and numerical results given in [15-16] where a hemispherical open-cavity receiver constructed with a coated copper tube through which thermal oil flows has been placed in the focal region of a 1.8 m outer diameter parabolic dish. The receiver tube forms the inner wall of the open-cavity receiver whose diameter is equal to the length (14 cm). One notice that, selective black chromium coating has been used on the copper tubes in order to minimize the thermal radiation losses (ε=0.1) and maximize the radiation absorption (α*=0.84). A complex and complete numerical model (including natural external convection, wind effect, internal forced convection, radiative and conductive heat losses) has been built by the authors who found mean deviation of 3.68%, regarding the thermal efficiency of the receiver by comparing their experimental and numerical results. We have chosen to compare the results of our simple model to the previous experimental results obtained at 13H30 to be to be sure that steady condition was reached. With the definitions given section 3 and the experimental data of [15-16], we found that parameter $\varepsilon = \frac{r^3}{q^4}$ was always lower than 2 10$^{-3}$. We have therefore calculated the fluid temperature by using equation (16) and found Tf=116°C which is to be compared to the reported experimental value of 135°C. Such a relative error (14%) show that the simple analytical model proposed in this work could be used in the early stage of solar receivers design. It is worth noting that our model exhibits one fluid temperature while in the experimental receiver, the fluid temperature evolves along the serpentine coils. The simple analytical model thus gives the exit temperature of the heating fluid.

## 5. Conclusion

A simple analytical model has been proposed for the calculation of the cavity and fluid temperatures of a solar cavity receiver placed in parabolic reflector dish. Approximate explicit solutions have also been deduced for small values of a parameter $\varepsilon$. Two examples of receivers (a large and a small one) have been considered. It has been shown that the approximate explicit solution is of sufficient accuracy when applied to calculate temperatures in the smallest cavity while the use of the complete analytical solution is necessary for the largest cavity. The model has also been compared to experimental results obtained in a hemispherical coiled cavity. It was found that the relative error in the exit temperature was of order of 14%. It is believed that these solutions could be useful in the early stage of solar receivers design as well as in teaching.